\documentclass[aps,prc,superscriptaddress,twoside,twocolumn,nofootinbib,showpacs]{revtex4-1}

\usepackage{amsmath,amssymb}
\usepackage{braket}
\usepackage{graphicx,bm}
\usepackage{tensor}
\usepackage{siunitx}

\usepackage{multirow,physics}

\begin{document}

\title{\boldmath
Decay angular distributions of $K^*$ and $D^*$ vector mesons in pion-nucleon scattering}


 \author{Sang-Ho \surname{Kim}}
 \email{sangho.kim@apctp.org}
 \affiliation{Asia Pacific Center for Theoretical Physics, Pohang, Gyeongbuk 37673, Korea}

\author{Yongseok \surname{Oh}}
\email{yohphy@knu.ac.kr}
\affiliation{Department of Physics, Kyungpook National University, Daegu 41566, Korea}
\affiliation{Asia Pacific Center for Theoretical Physics, Pohang, Gyeongbuk 37673, Korea}

\author{Alexander I. \surname{Titov}}
\email{atitov@theor.jinr.ru}
\affiliation {Bogoliubov Laboratory of Theoretical Physics, JINR, Dubna 141980, Russia}


\begin{abstract}
The production mechanisms of open strangeness ($K^*$) and open charm ($D^*$) vector mesons in 
$\pi^- p$ scattering, namely, $\pi^- +p \to K^{*0} + \Lambda$ and $\pi^- + p \to D^{*-} + \Lambda_c^+$,
are investigated within the modified quark-gluon string model.
In order to identify the major reaction mechanisms, we consider the subsequent decays of the
produced vector mesons into two pseudoscalar mesons, i.e., $K^* \to K + \pi$ and 
$D^* \to D + \pi$.
We found that the decay distributions and density matrix elements are sensitive to the production
mechanisms and can be used to disentangle the vector trajectory and pseudoscalar trajectory
exchange models.
Our results for $K^*$ production are compared with the currently available experimental data and
the predictions for $D^*$ production processes are presented as well.
Our predictions can be tested at the present or planned experimental facilities.
\end{abstract}

\pacs{13.85.-t, 
12.40.Nn,	
13.75.Gx,	
13.25.-k	
}

\maketitle

\section{Introduction}

Investigation of open charm and open strangeness production processes is one of major hadron 
physics programs at current or planned accelerator facilities which are supposed to provide pion 
beams~\cite{MNNS12} or antiproton beams~\cite{Friese06}.%
\footnote{The details of the physics programs using these beams can be found, for example,
in the websites of Japan Proton Accelerator Research Complex (J-PARC)~\cite{J-PARC} and
Facility for Antiproton and Ion Research in Europe (FAIR)~\cite{FAIR-GSI}.}
These facilities are expected to produce high-quality beams at energies high enough to produce
strange or charm hadrons.
Understanding the dynamics of charm and strange quarks is an interesting topic
as their mass scale is between the light quark sector which is dominated by chiral symmetry and the
heavy quark sector where heavy quark spin symmetry takes a crucial role.
Therefore, they are interpolating the two extreme regions and the deviations or corrections to chiral
symmetry and heavy quark symmetry are crucial to understand the underlying dynamics of the strong
interaction.

The heavy mass of the charm quark, in particular, leads to a very rich hadron spectrum with open or
hidden charm flavor, which includes exotic states that were not observed in the light quark sector.
Therefore, charm hadron spectroscopy is expected to open a new opportunity for unraveling
the strong interaction.
Besides, many interesting ideas using charm flavor have been suggested, which include the utilization 
of charm particles as a probe of nuclear medium at maximum compression, the study of the 
properties of exotic $XYZ$ mesons and so on~\cite{Swanson06,BEHV10}.

One of the important issues which are not fully resolved at present is the charm/strangeness production 
mechanism in hadron reactions.
Since the reaction energy is not high enough to be treated asymptotically, the widely-used models
for heavy quark production based on perturbative Quantum Chromodynamics (QCD) are not
applicable, and an essential improvement by including nonperturbative contributions is indispensable.
In the perturbative QCD approaches, in fact, charm quarks are produced through gluon fragmentations.
In order to produce charm quarks in peripheral collisions, however, such gluons must have a large
momentum (large $x \sim 1$), which is much larger than its average magnitude ($x \lesssim 0.2$)
inside a nucleon.
As a result, this mechanism can hardly be the major production mechanism for heavy flavor production
at relatively low energies.

Therefore, it is legitimate to rely on the approaches based on non-perturbative QCD
background for describing peripheral reactions.
In the present work, we adopt the quark-gluon string model (QGSM) developed by Kaidalov and
collaborators in Refs.~\cite{Kaidalov80,Kaidalov82,KP86}, which has been applied for the evaluation
of cross sections of the exclusive $\Lambda_c$ production in $pp$ and
$\bar{p} p$ collisions~\cite{KV94,TK08,KKMW11} and in $\pi p$
collisions~\cite{BK83,KHKN15,KKH16}.
A novel feature of this model is that the invariant amplitude of the binary reaction has a form of
the Regge amplitude, where the parameters of an effective ``Reggeons'' are determined by unitary
condition and additivity of intercepts and of inverse slopes of the Regge trajectories.

For such an effective Reggeons or effective meson-trajectories exchanges, usually only the vector 
meson exchange is considered.
Since the intercept of vector meson trajectory is larger than that of the corresponding pseudoscalar 
trajectory with the same flavor quantum number, the exchange of pseudoscalar meson trajectory is
expected to be suppressed.
Therefore, this model would be justified at large center-of-momentum energy squared $s$
and small magnitude of the squared momentum transfer $\abs{t}$.
However, in order to fully understand the production mechanisms, more physical quantities should be 
examined other than cross sections.
In particular, such physical quantities should be sensitive to the production mechanisms whose contribution
to cross sections is relatively small.
In fact, as we shall see below, the available data for $K^*$ production suggest that the vector meson trajectory 
exchange model needs to be modified to some extent.

In the present work, we elaborate on the angular distributions of pseudoscalar mesons originated from
the decays of vector mesons produced in $\pi N$ collisions.
More specifically, we consider the production of $K^*$ and $D^*$ vector mesons,
which decay into $K\pi$ and $D\pi$, respectively.
Therefore, the processes under consideration in the present work are the two-step reactions of 
$\pi N \to K^*\Lambda \to (K \pi) \Lambda$ and $\pi N \to D^*\Lambda_c \to (D \pi) \Lambda_c$, 
where we will specifically work on $\pi^- p$ collisions.
In particular, we focus on the angular distributions of $K$ and $D$ mesons produced by
these reactions, which bear the information on the production mechanisms of $K^*$ and
$D^*$ vector mesons.

This paper is organized as follows.
In Sec.~\ref{section:model} we describe QGSM, which will be used to describe $K^*$ and $D^*$
vector mesons.
All the theoretical tools to investigate the angular distributions of $K$ and $D$ mesons produced
by the decays of the corresponding vector mesons are detailed as well.
Then, in Sec.~\ref{section:results}, we show the results on cross sections, spin-density matrix elements,
and decay angular distributions of vector mesons produced in $\pi^- p$ collisions.
We summarize and conclude in Sec.~\ref{section:summary}.

\section{The model} \label{section:model}

The reactions under consideration in the present work are $\pi^- + p \to V + Y \to (P + \pi) + Y$, where 
$Y$, $V$, $P$ are flavored baryon, vector meson, and pseudoscalar meson, respectively.
In the strangeness sector, $Y = \Lambda(1116,1/2^+)$, $V = K^*(892,1^-)$,
and $P = K(494,0^-)$, while, in charm sector, $Y = \Lambda_c(2286,1/2^+)$, $V = D^*(2010,1^-)$,
and $P = D(1870,0^-)$~\cite{PDG16}.

The corresponding cross section for (2-body $\to$ 3-body) reactions reads
\begin{eqnarray}
d\sigma = \left( \frac{1}{16\pi\lambda_i} \abs{T_{fi}}^2 dt \right)
\times \left(\frac{k_fd\Omega_f dM_V}{16\pi^3} \right),
\label{EQ1}
\end{eqnarray}
where $T_{fi}$ is the invariant amplitude for the production process
and $\lambda_i \equiv \lambda (M_\pi^2,M_N^2,s)$ is the K\"all\'en function defined
as $\lambda (x,y,z) \equiv x^{2}+y^{2}+z^{2}-2xy-2yz-2zx $.
Here, $M_\pi$ and $M_N$ stand for the pion mass and the nucleon mass, respectively, and
we use $M_V$ for the vector meson mass.
The Mandelstam variables for the production process are defined as 
$s = (p_\pi^{} + p_p^{})^2 = (p_V^{} + p_Y^{})^2$
and $t = (p_p^{} - p_Y^{})^2 = (p_\pi^{} - p_V^{})^2$, where $p_\pi^{}$, $p_p^{}$, $p_V^{}$, and
$p_Y^{}$ are the four momenta of the pion, proton, produced (virtual) vector meson, and hyperon,
respectively.
The solid angle and the magnitude of the three momentum of outgoing pseudoscalar meson 
in the rest frame of the vector meson are represented by $\Omega_f$ and $k_f$, respectively.
The averaging over the initial spin states and sum over the final spin states is understood as well.

The invariant amplitude can be expressed as
\begin{eqnarray}
T_{fi} = \mathcal{A}_{m_f^{}, \lambda_V^{}; \,m_i^{}}
\frac{1}{p_V^2 - M_0^2 + i M_0 \Gamma_{\rm tot}}
\mathcal{D}_{\lambda_V^{}}(\Omega_f),
\label{EQ2}
\end{eqnarray}
where $m_i^{}$ and $m_f^{}$ denote the spin projections of incoming and outgoing baryons,
respectively, and $\lambda_V^{}$ represents the spin projection of the produced virtual vector meson.
$M_0$ and $\Gamma_{\rm tot}$ are the pole mass and the total decay width of the produced vector meson,
respectively.
The amplitudes of the $\pi^- + p \to V + Y$ and $V \to P + \pi$ reactions are denoted by
$\mathcal{A}$ and $\mathcal{D}$, respectively.
The decay process of vector mesons is considered in its rest frame.
In this case, the amplitude of the vector meson decay into two pseudoscalar mesons has the simple
form of
\begin{eqnarray}
\mathcal{D}_\lambda = 2c \sqrt{\frac{4\pi}{3}} Y_{1\lambda}(\Omega_f),
\label{EQ3}
\end{eqnarray}
where the constant $c$ is related to the $V \to P + \pi$ decay width $\Gamma_f^{}$ as
\begin{eqnarray}
c^2 = \frac{6\pi M_V^2\Gamma_f^{}}{k_f^{}} ,
\label{EQ4}
\end{eqnarray}
with $k_f$ being the magnitude of the three momentum of the final-state particles in the 
rest frame of the vector meson.
Integration of $d\sigma$ in Eq.~(\ref{EQ1}) over $dM_V$ and $d\Omega_f$ leads to the well-known
result for the corresponding unpolarized cross section,
\begin{eqnarray}
\frac{d\sigma}{dt} = \frac{\mbox{Br}}{16\pi\lambda_i} \abs{\mathcal{A}_{fi} }^2 ,
\label{EQ5}
\end{eqnarray}
with $\mbox{Br} = {\Gamma_f}/{\Gamma_{\rm tot}}$ when $\Gamma_{\rm tot} \ll M_V$.

\begin{figure*}[t]
\includegraphics[width=0.8\textwidth]{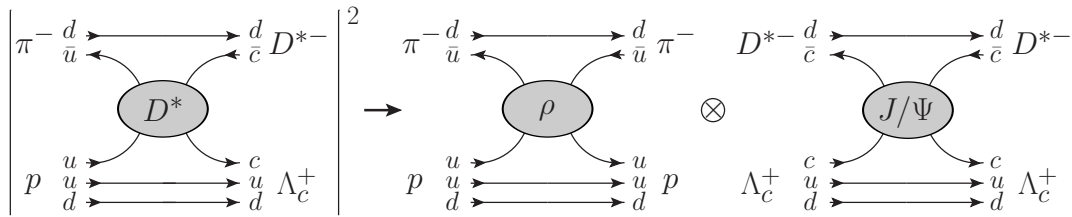}
\caption{Planar diagram decomposition for the reaction of $\pi^- + p \to D^{*-} + \Lambda_c^+$.}
\label{Fig:1}
\end{figure*}

Recent studies of strangeness and charm production at a few dozen GeV show that this cross section
can be successfully evaluated in the framework of QGSM suggested by
Kaidalov~\cite{Kaidalov80,Kaidalov82} and later developed and refined in a number of theoretical
works developed, for example, in Refs.~\cite{BK83,KP86,KV94,TK08,KKMW11,KHKN15,KKH16}.
QGSM is based on the planar quark diagram decomposition and unitary
conditions~\cite{Kaidalov82}, and it allows to represent the amplitude of the binary $a+b\to c+d$
reaction in terms of an effective Regge amplitude, where the effective trajectory
$\alpha_\mathcal{R}^{}(t)$ and the energy scale parameter $s_{ab;cd}$ are determined by
the well-established parameters of the elastic $a+b \to a+b$ and $c+d \to c+d$ reactions using
so-called the planar diagram decomposition.
An example of the planar diagram decomposition is depicted in Fig.~\ref{Fig:1} for the reaction
of $\pi^- + p \to D^{*-} + \Lambda_c^+$, where it is assumed that the amplitude is dominated by
the effective $D^*$ trajectory with parameters completely determined by the non-linear $\rho$ and
$J/\psi$ meson trajectories as found from the meson spectroscopy
studies~\cite{Kaidalov82,BGH97b,BBG00}.
Similarly, one can write the planar diagram decomposition for the $K^*\Lambda$ production with
substitution of the $J/\psi$ trajectory by the $\phi$ meson trajectory.
We refer the details to Ref.~\cite{TK08}.

\begin{figure*}[t]
\includegraphics[width=0.7\textwidth]{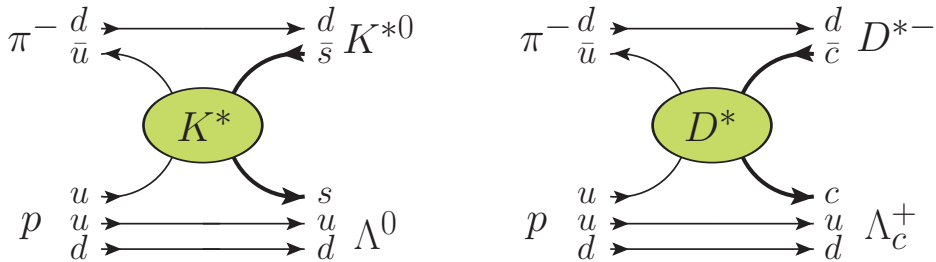}
\caption{
Diagrammatic representation of the effective $\pi^- + p \to K^{*0} + \Lambda$ and
$\pi^- + p \to D^{*-} + \Lambda_c^+$ reactions.}
\label{Fig:2}
\end{figure*}

Diagrammatic representations of the effective $\pi^- + p \to K^{*0} + \Lambda$ and
$\pi^- + p \to D^{*-} + \Lambda_c^+$ reactions are shown in Fig.~\ref{Fig:2}.
The corresponding spin-independent amplitudes read
\begin{eqnarray}
\mathcal{A}^V_{fi} = g_0^2 \, \frac{s}{\bar s} \,
\Gamma \bm{(}1 - \alpha^V_{\mathcal{R}}(t) \bm{)}
\left( \frac{s}{s^V_{0\mathcal{R}}} \right)^{\alpha^V_{\mathcal{R}}(t) - 1}
\label{EQ6}
\end{eqnarray}
with $\alpha^V_{{\cal R}_{p\Lambda}}(t) = 0.414 + {0.707}\, t $,
$s^V_{0{\cal R}_{p\Lambda}} = {1.66} \mbox{ GeV}^2$, $\bar s_{p\Lambda} = 1 \mbox{ GeV}^2$ for
$\pi^- + p \to K^{*0} + \Lambda$, and $\alpha^V_{{\cal R}_{p\Lambda_c}}(t)= -1.02 + {0.467}\, t $,
$s^V_{0{\cal R}_{p\Lambda_c}} = 4.75 \mbox{ GeV}^2$, $\bar s_{p\Lambda_c}=1$~GeV$^2$ for
$\pi^- + p \to D^{*-} + \Lambda_c^+$.
The trajectories of $\rho$, $\phi$, and $J/\psi$, as well as the energy-scale parameters
$s^V_{0\cal R}$ are determined following the prescription described in Ref.~\cite{TK08}.
The residual factor $g_0^{}$ will be determined in the next Section by comparison with the 
available experimental data for the $\pi^- + p \to K^{*0} + \Lambda$ reaction,
which leads to $g_0^2/4\pi \simeq 0.796$.

Since the angular distributions of pseudoscalar mesons produced through the decays of
$K^* \to K + \pi$ and $D^* \to D + \pi$ strongly depend on the spin of the participating particles,
the spin structure of the reaction amplitudes of Eq.~\eqref{EQ6} should be specified.
This, in fact, is the key component which can distinguish different production mechanisms.
It can be done by ``dressing" the spin-independent amplitude by the spin-factor $S_{fi}$
which carries the symmetry of exchanged Reggeon~\cite{TK08}, i.e.,
\begin{eqnarray}
\mathcal{A}_{fi} \to {\cal A}_{m_f^{},\lambda_V^{};\,m_i^{}} =
\mathcal{A}_{fi} \,\frac{1}{\mathcal{N}}\, S_{m_f^{},\lambda_V^{};\,m_i^{}},
\label{EQ7}
\end{eqnarray}
with the normalization factor
\begin{eqnarray}
\mathcal{N}^2 = \sum\limits_{m_f^{},m_i^{},\lambda_V^{}}  \abs{S_{m_f^{},\lambda_V^{};m_i^{}}}^2 .
\label{EQ8}
\end{eqnarray}
The $K^*$ meson couplings in the spin-factor $S_{fi}$ reads
\begin{widetext}
\begin{eqnarray}
S_{m_f^{},\lambda_V^{};m_i^{}} =
\epsilon^{\mu\nu\alpha\beta} q_{\mu}^{} p_{V\alpha}^{} \varepsilon^*_\beta (\lambda_V)
\times \bar u_{m_f^{}}(\Lambda)
\left[ (1+\kappa_{K^*p\Lambda}^{}) \gamma_\nu  - \kappa_{K^*p\Lambda}^{}
\frac{(p_p+p_\Lambda)_\nu}{M_p+M_\Lambda} \right] u_{m_i^{}}(p) ,
\label{EQ9}
\end{eqnarray}
\end{widetext}
where $q = p_V^{} - p_\pi^{} = p_p^{} - p_\Lambda^{}$ is the momentum transfer and
$\kappa_{K^*p\Lambda}^{} = 2.79$ is the tensor coupling constant
obtained from the average value of the Nijmegen soft-core potential~\cite{SR99,RSY99}.
The Dirac spinors of initial baryon and final baryon are denoted by $u_{m_i^{}}$ and $u_{m_f^{}}$, respectively,
and $\varepsilon (\lambda_V^{})$ is the polarization vector of the produced vector meson.
Generalization to the case of charm production may be achieved by the substitution
$M_\Lambda \to M_{\Lambda_c}$, $M_{K^*} \to M_{D^*}$, and so on.
Because of the lack of information, we assume $\kappa_{K^*p\Lambda}^{} = \kappa_{D^*p\Lambda_c}^{}$ 
as in Ref.~\cite{CERN-MPI-78}.
The normalization factor $\cal N$ in Eq.~(\ref{EQ7}) is introduced to compensate for the artificial $s$ and $t$ 
dependence generated by $S_{fi}$.

The differential cross section is then written as
\begin{eqnarray}
\frac{d\sigma}{dt \, d\Omega_f} = \frac{d\sigma}{dt}\,W(\Omega_f),
\label{EQ10}
\end{eqnarray}
where
\begin{eqnarray}
W(\Omega_f) &=& \sum\limits_{m_i^{},m_f^{},\lambda_V^{},\lambda'_V}
\mathcal{M}_{m_f^{},\lambda_V^{};m_i^{}}
\mathcal{M}_{m_f^{},\lambda'_V;m_i^{}}^*
\nonumber \\ && \mbox{} \times
Y_{1\lambda_V^{}}(\Omega_f)\,Y^*_{1\lambda'_V}(\Omega_f),
\label{EQ11}
\end{eqnarray}
with
\begin{eqnarray}
\mathcal{M}_{m_f^{},\lambda_V^{};m_i^{}} = \frac{1}{\mathcal{N}}\,
S_{m_f^{},\lambda_V^{};m_i^{}} .
\label{EQ12}
\end{eqnarray}
For definiteness with isospin quantum number we consider 
$K^{*0} \to K^+ \pi^-$ and $D^{*-} \to D^- \pi^0$ decays.
As is well known, since the decay angular distribution of outgoing $K^+$ or $D^-$ is analyzed 
in the virtual vector meson's rest frame, there is an ambiguity in choosing the quantization axis. 
One may choose the quantization axis anti-parallel to the outgoing hyperon $Y$ in
the center-of-momentum frame of the production process.
Or the quantization axis may be defined to be parallel to the incoming pion, i.e., the 
initial beam direction.
Following the convention of Ref.~\cite{CGLS72}, the former is called the $s$-frame and
the latter the $t$-frame.%
\footnote{In the case of vector meson photoproduction, the former is called the helicity 
frame, while the latter corresponds to the Gottfried-Jackson frame~\cite{SSW70}.}

The decay probabilities are expressed in terms of the spin-density matrix elements
$\rho_{\lambda\lambda^\prime}^{}$, where $\lambda_V^{}$ is abbreviated as $\lambda$,
which are determined by the amplitudes of Eq.~(\ref{EQ12}).
Depending on the polarization state of the initial and final states, we are interested in 
the following two cases:
\begin{enumerate}
\item
Unpolarized case, where the spin-density matrix is given by
\begin{eqnarray}
\rho^0_{\lambda\lambda'} = \sum\limits_{m_i^{} = \pm\frac12,\, m_f^{} = \pm\frac12}
\mathcal{M}_{m_f^{},\lambda^{};\,m_i^{}}\, \mathcal{M}^*_{m_f^{},\lambda';\,m_i^{}} ,
\label{EQ13}
\end{eqnarray}
\item
Recoil polarization case, when the spin of the outgoing hyperon ($Y$) is determined 
by their decay distribution using that it is self-analyzing.
Then, depending on the spin state of the hyperon, we have two kinds of spin-density 
matrices defined as
\begin{eqnarray}
\rho^{\pm}_{\lambda\lambda'} = \sum\limits_{m_i^{} = \pm\frac12}
\mathcal{M}_{m_f^{},\lambda;m_i^{}}\, \mathcal{M}^*_{m_f^{},\lambda';m_i^{}} .
\label{EQ14}
\end{eqnarray}
Here, $\rho^+$ and $\rho^-$ correspond to the cases when the spin or helicity of the produced hyperon  
is $m_f^{} = + \frac12$ and $-\frac12$, respectively.
\end{enumerate}

Denoting the polar and the azimuthal angles of the outgoing pseudoscalar $K$ (or $D$) mesons
by $\Theta$ and $\Phi$, respectively, the decay angular distributions can be expressed
in terms of the spin-density matrix elements as
\begin{eqnarray}
W^{0}(\Omega_f) &=& \frac{3}{4\pi} \Bigl[ \rho^0_{00}\cos^2\Theta + \rho^0_{11}\sin^2\Theta
- \rho^0_{1-1} \sin^2\Theta \cos2\Phi
\nonumber\\ && \mbox{}
- {\sqrt{2}}\, \mathrm{Re}(\rho^0_{10}) \sin2\Theta\cos\Phi \Bigr],
\label{EQ15}
\end{eqnarray}
for the unpolarized case and
\begin{eqnarray}
W^{\pm}(\Omega_f) &=& \frac{3}{4\pi} \biggl[\rho^{\pm}_{00}\cos^2\Theta +
\frac12\left(\rho^{\pm}_{11} + \rho^{\pm}_{-1-1}\right)\sin^2\Theta
\nonumber\\ && \mbox{}
- \rho^{\pm}_{1-1}\sin^2\Theta\cos2\Phi
\nonumber \\ && \mbox{}
- \frac{1}{\sqrt{2}}
\mathrm{Re}\left(\rho^{\pm}_{10}-\rho^{\pm}_{-10}\right)
\sin2\Theta\cos\Phi \biggr],
\label{EQ16}
\end{eqnarray}
for the case of recoil polarization.
Here, we made use of the Hermitian conditions: $\rho_{-11}^{} = \rho_{1-1}^{}$,
$\rho_{01}^{} = \rho_{10}^{}$, and $\rho_{0-1}^{} = \rho_{-10}^{}$.
In addition, for unpolarized reaction, we also have the sum rule
$\rho^0_{00}+\rho^0_{11}+\rho^0_{-1-1}=1$ and the symmetry conditions,
$\rho^0_{11}=\rho^0_{-1-1}$ and $\rho^0_{01}=-\rho^0_{0-1}$.
In the case of recoil polarization, however, these additional relations do not hold.

As was mentioned earlier, the purpose of the present work is to test the validity of the 
dominance of vector meson trajectory exchange.
This assumption is based on the observation that the intercept of the $K^*$ ($D^*$) vector meson, for 
instance, is larger than that of the corresponding pseudoscalar $K$ ($D$) 
meson trajectory~\cite{BGH97b}.
However, other mechanisms cannot be excluded, and the contribution from such mechanisms
should be verified by physical quantities related to the spin structure of the production mechanisms.
In fact, as we will see later, the available data for density matrix elements suggest that there exist
contributions from mechanisms other than vector trajectory exchange.
Therefore, in addition to vector trajectory exchanges, we consider the exchanges of effective 
pseudoscalar $K$ and $D$ trajectories.
In this case, the spin-independent amplitude reads
\begin{eqnarray}
\mathcal{A}^{PS}_{fi} \simeq g_0^2 \,
\Gamma \bm{(} -\alpha^{\rm PS}_{\mathcal{R}}(t) \bm{)}
\left( \frac{s}{s^{\rm PS}_{0\mathcal{R}}} \right)^{\alpha^{\rm PS}_{\mathcal{R}}(t)}
\label{EQ17}
\end{eqnarray}
with $\alpha^{\rm PS}_{\mathcal{R}_{p\Lambda}}(t) = -0.151 + {0.617}\, t $,
$\alpha^{\rm PS}_{\mathcal{R}_{p\Lambda_c}}(t) =  {-1.611} + {0.439}\, t$~\cite{BBG00}.
The energy scale parameters determined by the flavor content of the vertices are assumed 
to be the same as in the vector meson exchange case so that
$s^{\rm PS}_{0\mathcal{R}_{p\Lambda}} = s^V_{0\mathcal{R}_{p\Lambda}}$ and
$s^{\rm PS}_{0\mathcal{R}_{p\Lambda_c}} = s^V_{0\mathcal{R}_{p\Lambda_c}}$.
The spin factor $S_{fi}$ now reads
\begin{eqnarray}
S^{\rm PS}_{m_f^{},\lambda_V^{};m_i^{}} = \varepsilon^{*}_\mu(\lambda_V)\,q^\mu
\bar u_{m_f^{}}(\Lambda) \gamma_5 u_{m_i^{}}(p).
\label{EQ18}
\end{eqnarray}

\begin{figure*}[t]
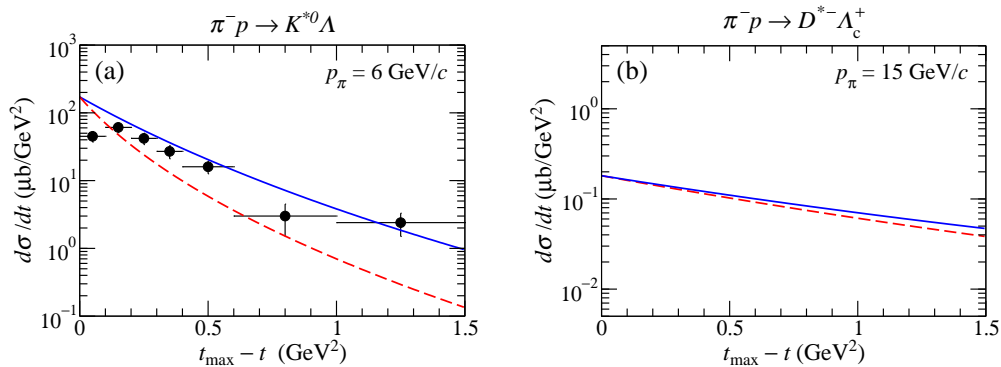

\includegraphics[width=0.35\textwidth]{fig3a.eps}\qquad
\includegraphics[width=0.35\textwidth]{fig3b.eps}
\caption{
Unpolarized differential cross sections of (a) $\pi^- + p \to K^{*0} + \Lambda$ and
(b) $\pi^- + p \to D^{*-} + \Lambda_c^+$ for the vector (solid curves) and pseudoscalar
(dashed curves) Reggeon exchanges.
The experimental data for $\pi^- + p \to K^{*0} + \Lambda$ are from Ref.~\cite{CGLS72}.}
\label{Fig:3}
\end{figure*}

\section{Results and Discussion} \label{section:results}

In this Section, we present numerical results on differential cross sections, spin-density
matrix elements, and decay angular distributions of $K$ and $D$ mesons in $\pi N$ scattering.

\subsection{Unpolarized cross sections}

By collecting all information, the unpolarized differential cross sections of the
$\pi^- + p \to K^{*0} + \Lambda$ and $\pi^- + p \to D^{*-} + \Lambda_c$ reactions
for the vector (V) and pseudoscalar (PS) effective Reggeon exchanges are written as
\begin{eqnarray}
\frac{d\sigma^{(\rm V)}}{dt} &=&
\frac{\pi}{\lambda_i} \left(\frac{s}{\bar s}\right)^2
\left[ \frac{(g^{\rm V}_0)^2}{4\pi} \right]^2
\left[ \Gamma \bm{(}1-\alpha^{\rm V}(t) \bm{)} \right]^2
\nonumber \\ && \mbox{} \times
\left(\frac{s}{s_{0\mathcal{R}^{\rm V}}} \right)^{2(\alpha^{\rm V}(t)-1)},
\nonumber\\
\frac{d\sigma^{(\rm PS)}}{dt}&=&
\frac{\pi}{\lambda_i} \left[ \frac{(g^{\rm PS}_0)^2}{4\pi}\right]^2
\left[ \Gamma \bm{(} -\alpha^{\rm PS}(t) \bm{)} \right]^2
\left(\frac{s}{s_{0\mathcal{R}^{\rm PS}}} \right)^{2 \alpha^{\rm PS}(t)}.
\nonumber \\
\label{EQ188}
\end{eqnarray}
The residual factor $g_0^2$ is, in general, a function of $t$, and should be determined 
by the comparison with experimental data.
We use ${(g^{\rm V}_0)^2}/{4\pi}=0.796$ for the vector meson trajectory
exchange, which is found from comparison with the available experimental data for $K^{*0}$ production.
We use this value for both the strangeness and charm production processes as we do not have any
data for charm vector meson production.
Since we are interested in identifying the major production mechanisms, we need to be able 
to distinguish between the pseudoscalar meson trajectory exchange and the vector meson 
trajectory exchange through measurable physical quantities.
Since the pseudoscalar exchange mechanism is expected to be small, we consider two
extreme cases, namely, vector-exchange dominance and pseudoscalar-exchange dominance.
For this purpose, we adjust the value of $g^{\rm PS}_0$ to achieve the condition that
$d\sigma^{(\rm PS)}/dt = d\sigma^{(\rm V)}/dt$ at zero vector meson production angle, i.e., at
$t = t_{\rm max}$.
This leads to ${(g^{\rm PS}_0)^2}/{4\pi}=1.1$ and 13.5 for the $K^*$ and $D^*$ mesons 
production, respectively.
Of course, the realistic case is between these two extreme cases, and the relative strength of
the two mechanisms should be determined by experimental data.

The obtained differential cross sections for $K^*$ and $D^*$ production are exhibited in
Figs.~\ref{Fig:3}(a) and (b), respectively.
Throughout the present study, the initial pion momentum in the laboratory frame is chosen
to be $p_\pi^{} = 6$~GeV/$c$ for strangeness production and 15~GeV/$c$ for charm production.
The vector (V) and pseudoscalar (PS) Reggeon exchanges are shown by the solid and dashed
curves, respectively, together with available experimental data of Ref.~\cite{CGLS72} for 
$K^*$ production.
Although the energy scale is different, it turns out that the cross section of charm 
production is suppressed compared with strangeness production, which is consistent with 
the observation made in Ref.~\cite{KHKN15}.
One can see that both the vector-type Reggeon exchange and the pseudoscalar-type Reggeon
exchange exhibit a similar $t$-dependence in differential cross sections.
This resemblance is clearly seen in the case of charm production, although the available data 
seem to prefer the vector-type exchange in the case of strangeness production.
Therefore, the $t$-dependence of cross sections cannot clearly distinguish the two exchanges.
As we will see in the next subsections, however, the situation changes for spin-density 
matrix elements and the angular distributions of
$K^*\to K\pi$ and $D^*\to D \pi$ decays, where the difference between the two types of
exchanges is revealed even at the qualitative level.

\subsection{Spin-density matrix elements}
\begin{figure*}[t]
\includegraphics[width=0.95\textwidth]{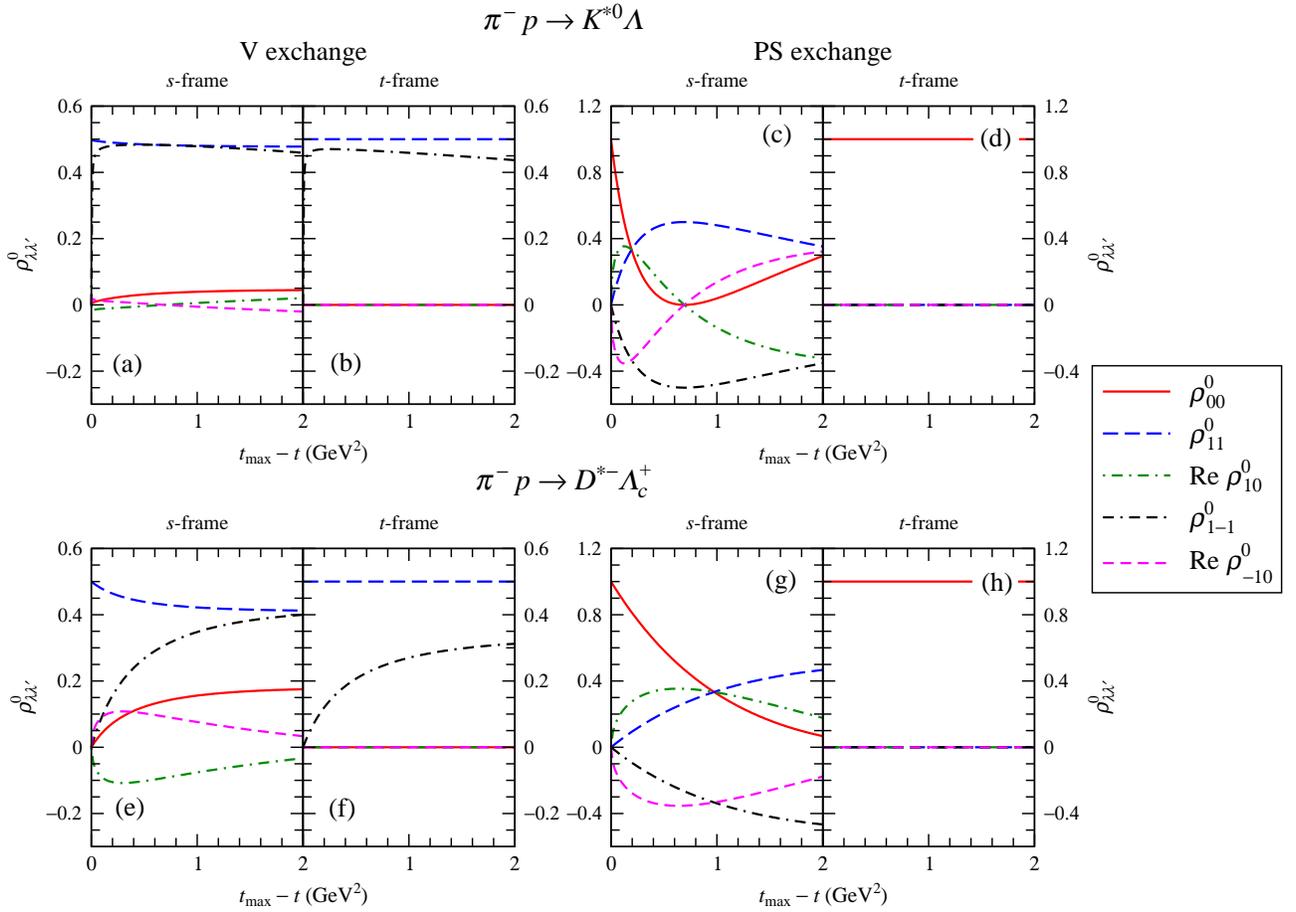}
\caption{The spin-density matrix elements $\rho^0_{\lambda\lambda'}$ defined in Eq.~(\ref{EQ13}) as
functions of $(t_{\rm max} - t)$ (a)--(d) for $K^{*-}$ production at $p_\pi^{} = 6$~GeV/$c$ and (e)--(f) for $D^{*-}$ 
production at $p_\pi^{} = 15$~GeV/$c$.
The results for vector meson (V) and pseudoscalar (PS) Reggeon exchanges are in (a), (b), (e), (f)
and (c), (d), (g), (h) panels, respectively.
The results in (a), (c), (e), (g) are obtained in the $s$-frame, while those in (b), (d), (f), (h) are
in the $t$-frame.}
\label{Fig:4}
\end{figure*}

The results for spin-density matrix elements $\rho^0_{\lambda\lambda'}$ defined in 
Eq.~(\ref{EQ13}) are presented in Fig.~\ref{Fig:4} for $K^{*0}$  and $D^{*-}$ production 
as functions of $(t_{\rm max} - t)$.
We also limit our consideration to relatively small values of $\abs{t}$ such that $\abs{t_{\rm max}-t} \leq 2$ 
GeV$^2$, where the applicability of QGSM can be justified.
Shown in Fig.~\ref{Fig:4} are the results for the vector-type Reggeon exchange model and for
the pseudoscalar-type Reggeon exchange model, which are calculated in the $s$- and $t$-frames.
Our results numerically confirm the symmetry properties,
$\rho^0_{11} = \rho^0_{-1-1}$,  $\rho^0_{\pm1 0}=\rho^0_{0\pm 1}$,
$\rho^0_{\pm 10}=-\rho^0_{0\mp 1}$, and $\rho^0_{1-1}=\rho^0_{-11}$.

\begin{figure*}[t]
\includegraphics[width=0.95\textwidth]{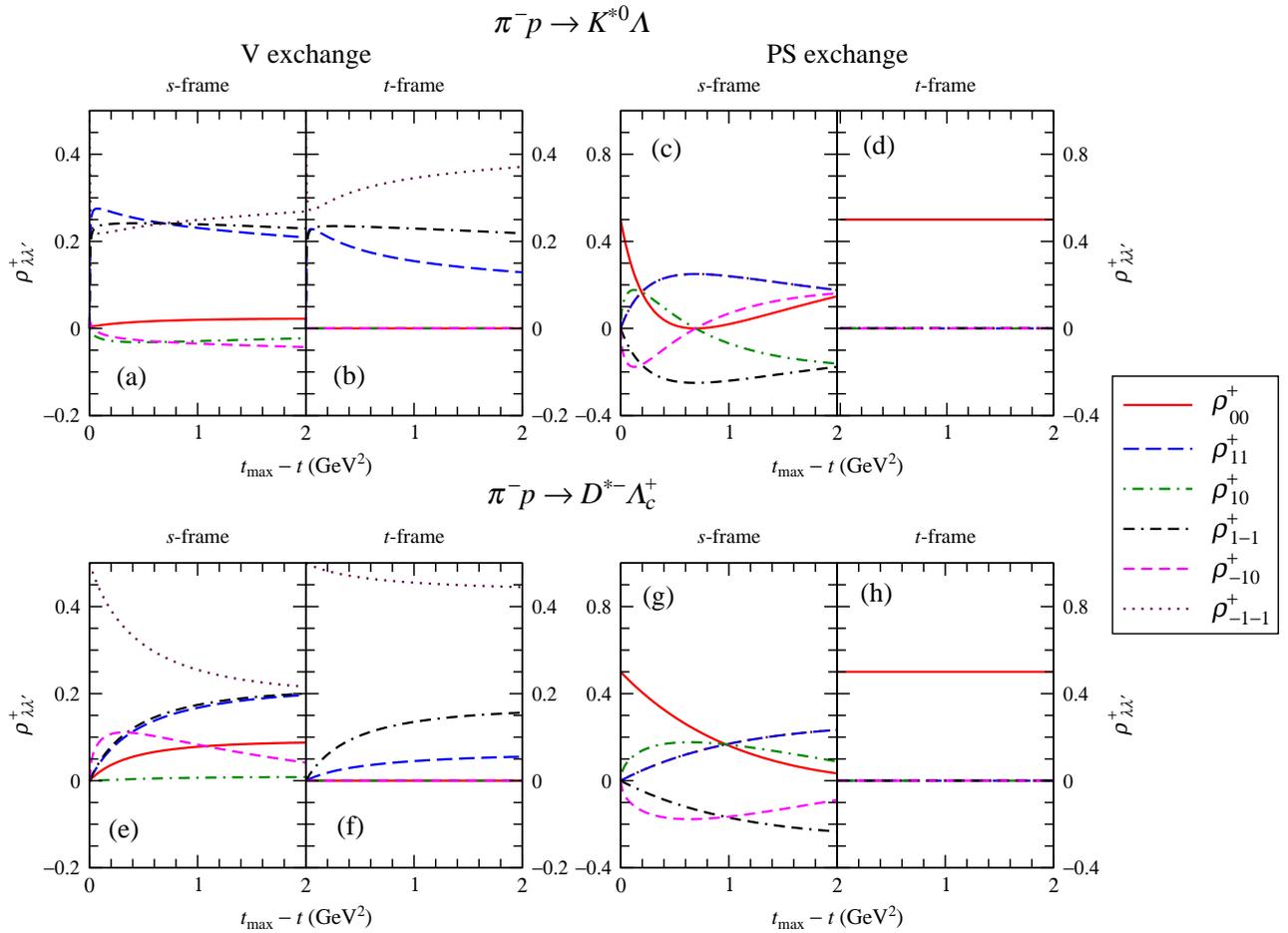}
\caption{
The same as in Fig.~\ref{Fig:4} but for $\rho^+_{\lambda\lambda'}$.}
\label{Fig:5}
\end{figure*}

In the case of vector-type Reggeon exchange, the matrix elements $\rho^0_{\lambda\lambda'}$
with $\abs{\lambda} = \abs{\lambda'} = 1$ are enhanced.
This ascribes to the spin structure 
$\epsilon^{\mu\nu\alpha\beta} q_{\mu}^{} p_{V\alpha}^{} \varepsilon^{*}_\beta(\lambda_V^{})$
of the amplitude in Eq.~(\ref{EQ9}).
In the vector meson rest frame, where $p_V^{} = (M_V^{},0,0,0)$ and $\bm{q} = - \bm{p}_\pi^{}$,
this factor is proportional to the vector product of ${\bm{\varepsilon}}^*(\lambda_V^{}) \times \bm{p}_\pi^{}$.
In the $s$-frame and small momentum transfers, $\bm{p}_\pi^{}$ has a large $z$ component 
and a small $x$ component, which leads to ${\bm\varepsilon}^*(\lambda_V^{}) \times \bm{p}_\pi^{}
\simeq i\lambda_V {\bm\varepsilon}^*(\lambda_V^{})|\bm{p}_\pi|$
and thus causes the large enhancement of $\rho^0_{\abs{\lambda} = 1,\, \abs{\lambda^\prime} = 1}$.
In the $t$-frame, $\bm{p}_\pi^{}$ is parallel to the quantization axis, and this leads to that
$\rho^0_{\lambda\lambda'}$ with either $\lambda = 0$ or $\lambda' = 0$ vanish.
We also note that $\rho_{1-1}^{0} = 0$ at $t = t_{\mathrm{max}}$.
This is because of the relation $\rho_{1-1}^{0} \propto \sin^2\theta$, where $\theta$ is the scattering 
angle of the vector meson in the c.m. frame for the scattering process.
So is the matrix element $\rho_{1-1}^{+}$ as seen in Fig.~\ref{Fig:5}.

In the case of pseudoscalar-type Reggeon exchange, however, the situation is quite different.
The production amplitude of this mechanism is proportional to the scalar product,
$\bm{\varepsilon}^*(\lambda_V^{}) \cdot \bm{p}_\pi^{}$, which leads to a strong enhancement of
$\rho^0_{00}$ in the $t$-frame, so that $\rho^0_{00}=1$ and all the other
$\rho^0_{\lambda\lambda'}$ vanish.

Shown in Fig.~\ref{Fig:5} are the results for $\rho^+_{\lambda\lambda'}$ defined in Eq.~(\ref{EQ14}).
In this case, the spin alignment of the outgoing hyperon is fixed to be $m_f^{} = +\frac12$.
The absolute values of $\rho^\pm$ are smaller than those of $\rho^0$ by about a factor of 
two because of the difference in the numerators in Eqs.~(\ref{EQ13}) and (\ref{EQ14}).
The spin-density matrix elements $\rho^-_{\lambda\lambda'}$ can be obtained  from
$\rho^+_{\lambda\lambda'}$ using the symmetry relations~\cite{SSW70}:
$\rho^-_{11}= \rho^+_{-1-1}$, $\rho^-_{00}= \rho^+_{00}$, $\rho^-_{-11}= \rho^+_{-11}$,
$\rho^-_{10}=-\rho^+_{0-1}$, and so on.

\begin{figure*}[t]
\includegraphics[width=0.7\textwidth]{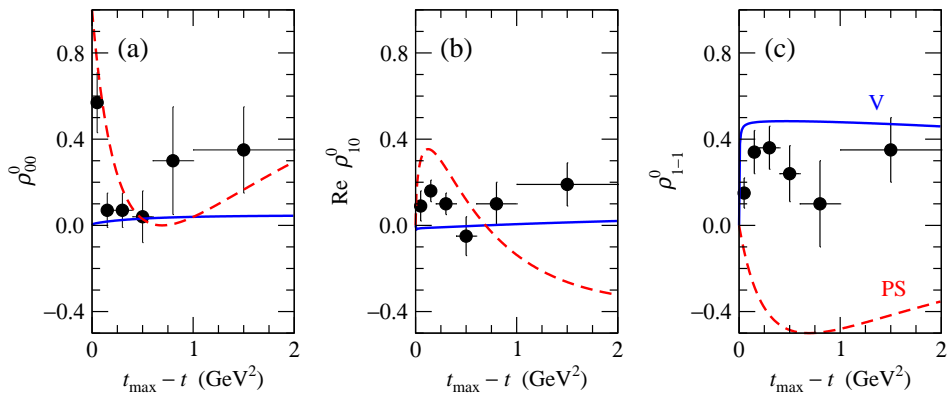}
\caption{
Spin-density matrix elements for $K^{*0}$ production in the $s$-frame.
The panels (a), (b), and (c) correspond to $\rho^0_{00}$, Re$\rho^0_{10}$, and $\rho^0_{1-1}$
matrix elements, respectively.
The vector and pseudoscalar Reggeon exchange models are depicted by the solid and dashed curves,
respectively.
The experimental data are from~\cite{CGLS72}.}
\label{Fig:6}
\end{figure*}

\begin{figure*}[t]
\includegraphics[width=0.7\textwidth]{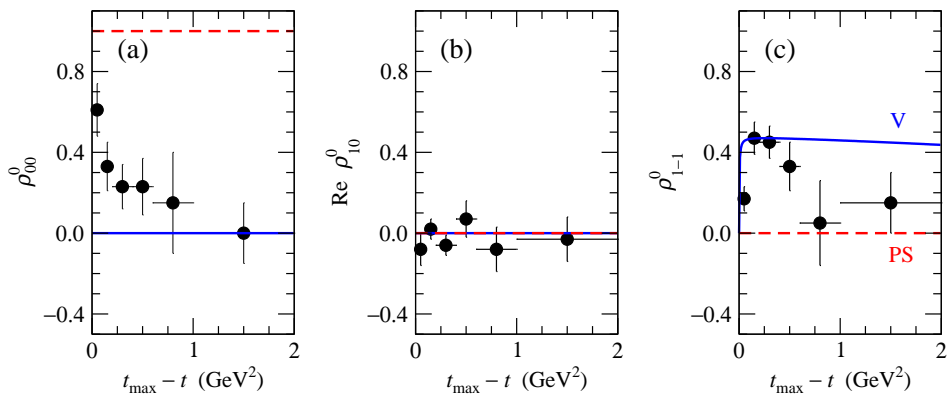}
\caption{
The same as in Fig.~\ref{Fig:6} but in the $t$-frame.}
\label{Fig:7}
\end{figure*}

In Figs.~\ref{Fig:6} and \ref{Fig:7}, we compare our results with the available
experimental data of Ref.~\cite{CGLS72} for $K^{*0}$ production at $s$- and $t$-frames, 
respectively.
Although the vector-exchange mechanism leads to a better agreement with the
data than the pseudoscalar-exchange model, we can see that the vector-exchange model
alone cannot successfully explain the data.%
\footnote{We could confirm this conclusion through the comparison with the data obtained at 
$p_{\pi}^{}=4.5$~GeV/c~\cite{CGLS72} as well.}
New experimental data for $K^*$ production with higher precision are, therefore,
strongly desired.
In $D^*$ production, the difference is also large enough to be verified by experiments and
the analyses can be done at current or future experimental facilities.

\subsection{Angular distributions of vector meson decays}

\begin{figure*}[t]
\includegraphics[width=0.7\textwidth]{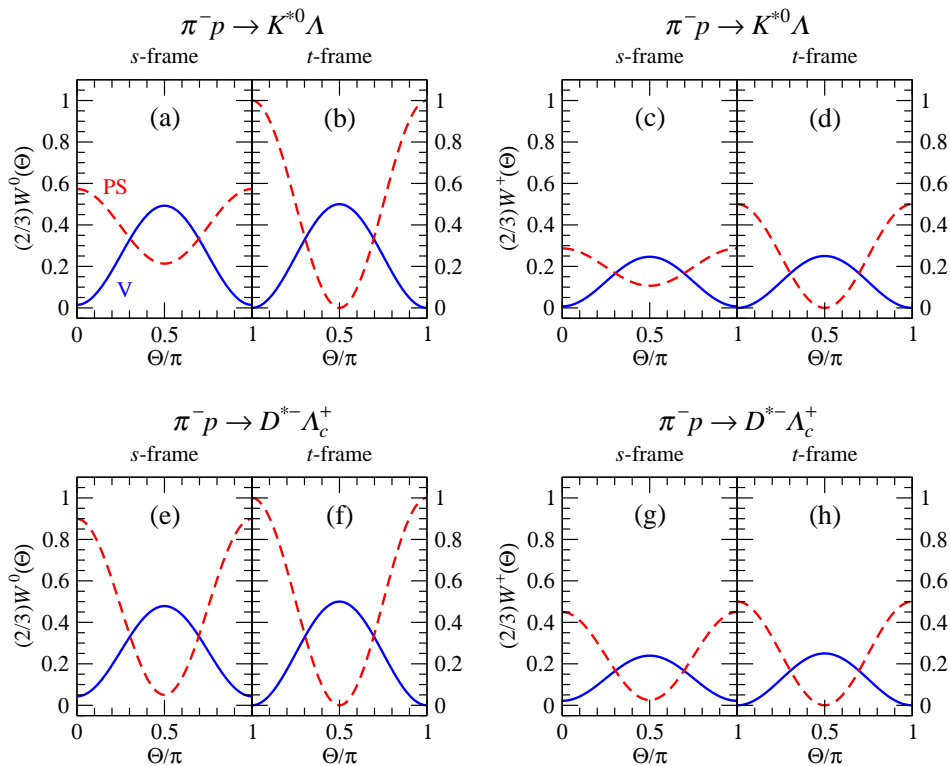}
\caption{
Angular distributions $\frac23\, W(\Theta)$ of Eq.~(\ref{EQ155}) for $K^*$ and $D^*$
excitations are shown in the upper and lower panels, respectively.
The panels (a), (b), (e), (f) are for $\frac23\,W^0(\Theta)$ and (c), (d), (g), (h) are for
$\frac23\,W^+(\Theta)$.
The results are given at both the $s$- and $t$-frames.
The vector and pseudoscalar Reggeon exchange cases are depicted by the solid and dashed curves,
respectively.
Calculation is done for $\abs{t_{\rm max}-t} = 0.1 \mbox{ GeV}^2$
at $p_\pi^{} = 6$~GeV/$c$ for $K^*$ production and $15$~GeV/$c$ for $D^*$ production.}
\label{Fig:8}
\end{figure*}

The polar angle distributions of outgoing $K$ and $D$ mesons are obtained by integrating
$W(\Theta,\,\Phi)$ of Eqs.~(\ref{EQ15}) and~(\ref{EQ16}) over the azimuthal angle $\Phi$, 
which gives
\begin{eqnarray}
\frac{2}{3}\,W^{0}(\Theta) &=& \rho^0_{00}\, \cos^2\Theta + \rho^0_{11}\, \sin^2\Theta ,
\nonumber\\
\frac{2}{3}\,W^{\pm}(\Theta) &=& \rho^{\pm}_{00}\, \cos^2\Theta
+ \frac12\left(\rho^{\pm}_{11}+\rho^{\pm}_{-1-1}\right)\sin^2\Theta.
\label{EQ155}
\end{eqnarray}
These distributions are presented in Fig.~\ref{Fig:8} for the production and decays of
$K^*$ and $D^*$ mesons at $\abs{t_{\rm max}-t} =  0.1 \mbox{ GeV}^2$ with $p_\pi^{} = 6$
and $15$~GeV/$c$, respectively.

In all cases, one can observe maxima at $\Theta=\frac{\pi}{2}$ for the vector trajectory exchange
while minima are observed at the same angle for the pseudoscalar trajectory exchange.
In other words, the distribution functions for the vector trajectory exchange display a cosine function
shape, while those of the pseudoscalar trajectory exchange show a sine function shape.
This is a direct consequence of the spin density matrix elements $\rho^0_{00}$ and
$\rho^0_{11}$ shown in Figs.~\ref{Fig:4} and \ref{Fig:5}.

\begin{figure*}[t]
\includegraphics[width=0.7\textwidth]{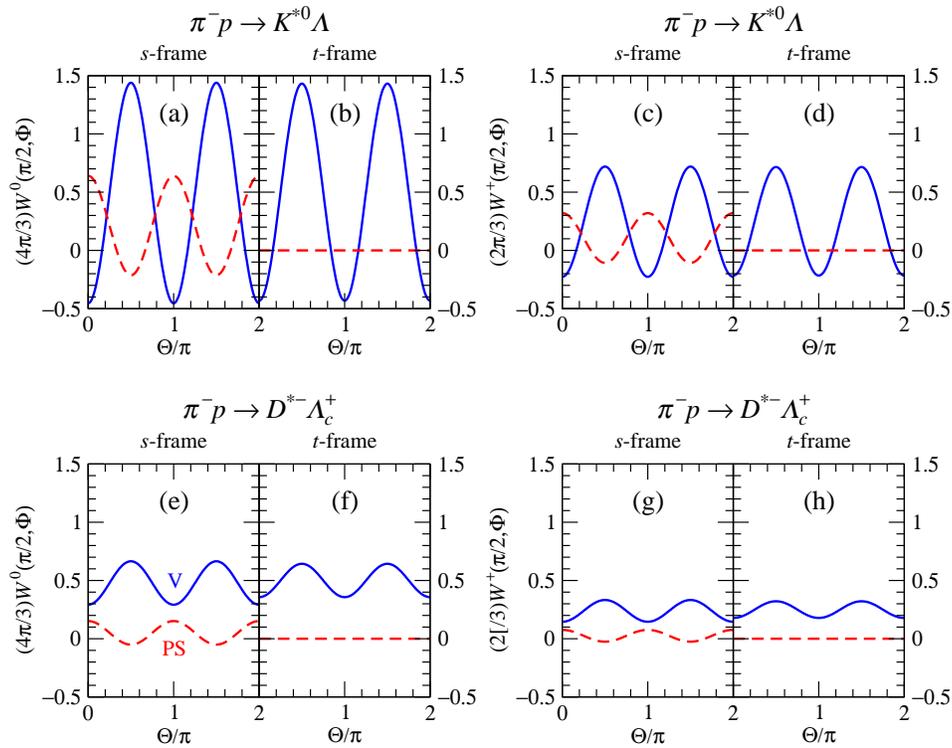}
\caption{
Same as in Fig.~\ref{Fig:8} but for the azimuthal angle distributions
$\frac{4\pi}{3}\,W(\Theta=\pi/2,\Phi)$ of Eq.~(\ref{EQ1555}).}
\label{Fig:9}
\end{figure*}

The azimuthal angle distributions at a fixed polar angle $\Theta$ can also be obtained from
Eqs.~(\ref{EQ15}) and (\ref{EQ16}).
At $\Theta=\frac{\pi}{2}$, we have
\begin{eqnarray}
\frac{4\pi}{3} W^{0}(\Theta=\frac{\pi}{2},\,\Phi) &=&
\rho^0_{11}-\rho^0_{1-1}\cos2\Phi~,
\nonumber\\
\frac{4\pi}{3} W^{\pm}(\Theta=\frac{\pi}{2},\,\Phi) &=&
\frac12\left(\rho^{\pm}_{11} + \rho^{\pm}_{-1-1}\right) 
\nonumber \\ && \mbox{}
- \rho^{\pm}_{1-1} \cos2\Phi.
\label{EQ1555}
\end{eqnarray}
The corresponding distributions are shown in Fig.~\ref{Fig:9}
at $\abs{t_{\rm max}-t} = 0.1 \mbox{ GeV}^2$.
In the $s$-frame, the matrix element $\rho^0_{1-1}$ takes a positive value for vector-type 
exchange and a negative value for pseudoscalar-type exchange.
This difference makes that $W(\frac{\pi}{2},\Phi)$ of vector-type exchange and 
pseudoscalar-type exchange are out of phase.
The amplitudes of the oscillations in $W^0$ are found to be larger than those of $W^\pm$,
which reflects the differences in $\rho_{1-1}^0$ as shown in Figs.~\ref{Fig:4} and \ref{Fig:5}.
For the pseudoscalar-Reggeon exchange in the $t$-frame, $\rho_{\lambda,\lambda'}^{0,+} = 0$ for
$\abs{\lambda} = \abs{\lambda'} = 1$, and, therefore, the corresponding distributions
$W(\frac{\pi}{2},\Phi)$ vanish identically.

\section{Summary and Conclusion} \label{section:summary}

In summary,  we investigated the reactions of open strangeness $K^*$ and open charm $D^*$ 
vector mesons in $\pi N$ scattering based on the quark-gluon string model.
We found that unpolarized cross sections of $K^*$ meson production is satisfactorily described 
by QGSM with vector trajectory exchange.
Although the contribution from pseudoscalar trajectory exchange is expected to be small, it 
also gives a similar $t$-dependence of differential cross sections as the vector-type exchange 
model. 
Therefore, differential cross sections cannot be used to disentangle the two production mechanisms.

In order to verify the mechanisms of vector meson production, we then studied the angular 
distributions of vector meson decays.
Unlike the cross sections, the spin density matrix elements are sensitive to the spin 
structure of the production amplitude and, as a result, they show very different $t$-dependence
and can be used to distinguish the vector and pseudoscalar exchanges.
Furthermore, the density distribution functions are found to have complete different angle-dependence 
depending on the production mechanisms and can be used to probe the spin structure of the 
reaction amplitudes.
In fact, the available data for spin density matrix elements of $K^*$ production show that 
the major production mechanism would be the vector-type exchange but it requires a noticeable 
contributions from the pseudoscalar-type exchange.
Because of the limited experimental data, we cannot estimate the relative strength between the 
vector and pseudoscalar exchanges, and, therefore, new data are strongly called for to 
investigate strangeness and charm production mechanisms.

We also presented our predictions for charm production.
In this case, the $t$-dependence of differential cross sections of the vector and pseudoscalar 
exchanges is even closer to each other because of the similarity in the slope of vector and 
pseudoscalar trajectories,
and thus the measurements of differential cross sections does not help pin down the production
mechanisms.
However, spin density matrix elements and decay angular distributions are very sensitive
to the production mechanisms as in the case of $K^*$ production, and more detailed studies on 
these quantities are expected to shed light on our understanding of the strong interaction.
In particular, the measurements for $K^*$ and $D^*$ productions are complementary to each other
and would be important to understand the dependence of the production mechanisms on the 
quark mass scale.
All these predictions can be tested and verified in future experimental programs with pion beams, for 
instance, at J-PARC facility.

\section*{Acknowledgments} 
A.I.T. thanks the Asia Pacific Center for Theoretical Physics (APCTP) for kind hospitality
given to him during his visit, which initiated this work.
The work of  Y.O. was supported by the National Research Foundation of Korea 
under Grant No. NRF-2015R1D1A1A01059603.

\end{document}